\documentclass[prl,superscriptaddress,showpacs,twocolumn]{revtex4}

\usepackage{graphicx}
\usepackage{amssymb,amsmath}

\usepackage[dvips]{color}

\newcommand{\eref}[1]{equation~(\ref{#1})}
\newcommand{\fref}[1]{figure~\ref{#1}}
\newcommand{\Fref}[1]{Figure~\ref{#1}}

\def\vk{{\bf k}}
\newcommand{\svek}{%
        \mathbf}
\newcommand{\dif}{%
        \hbox{d}}
\newcommand{\tr}{%
        \hbox{ tr}}

\def\delomc{\delta\omega_c}

\newcommand{\fermi}[1]{%
	\hbox{f($#1$)}}

\def\beq{\begin{equation}}
\def\eeq{\end{equation}}
\def\vk{{\bf k}}

\newcommand{\im}{%
           \imath}

\begin{document}

\title{Rare-Earth vs. Heavy Metal Pigments\\ and their Colors from First Principles}

\author{Jan M. Tomczak}
\affiliation{Department of Physics and Astronomy, Rutgers University, Piscataway, New Jersey 08854, USA}
\author{L. V. Pourovskii}
\affiliation{Centre de Physique Th{\'e}orique, Ecole Polytechnique, CNRS UMR7644, 91128 Palaiseau, France}
\author{L. Vaugier}
\affiliation{Centre de Physique Th{\'e}orique, Ecole Polytechnique, CNRS UMR7644, 91128 Palaiseau, France}
\author{A. Georges}
\affiliation{Centre de Physique Th{\'e}orique, Ecole Polytechnique, CNRS UMR7644, 91128 Palaiseau, France}
\affiliation{Coll{\`e}ge de France, 11 place Marcelin Berthelot, 75005 Paris, France}
\affiliation{DPMC-MaNEP, Universit\'e de Gen\`eve, 24 quai Ernest Ansermet, CH-1211 Gen\`eve, Suisse}
\affiliation{Japan Science and Technology Agency, CREST, Kawaguchi 332-0012, Japan}
\author{S. Biermann}
\affiliation{Centre de Physique Th{\'e}orique, Ecole Polytechnique, CNRS UMR7644, 91128 Palaiseau, France}
\affiliation{Japan Science and Technology Agency, CREST, Kawaguchi 332-0012, Japan}

%


\maketitle

%
{\bf 
Many inorganic pigments contain heavy metals  
hazardous to health and environment. 
Much attention has been devoted to 
the quest for 
non-toxic alternatives 
based on rare-earth elements. 
The computation of colors from first principles is a challenge to electronic 
structure methods however, especially for 
materials with localized $f$-orbitals. 
Here, starting from atomic positions only, we compute the 
color of the 
red pigment cerium fluorosulfide CeSF, as well as of mercury sulfide 
HgS (classic `vermilion'). 
Our methodology employs many-body theories 
to compute the optical absorption, combined with an intermediate length-scale modelization 
to assess how coloration depends on film thickness, pigment concentration and granularity.
We introduce a quantitative criterion for the performance of a pigment. 
While for HgS this criterion is satisfied due to large 
transition matrix elements between wide bands, CeSF presents an alternative paradigm: 
the bright red color is shown to stem from the combined effect 
of the quasi two-dimensionality and the localized nature of $4f$ states.  
Our work demonstrates the power of modern computational methods, 
with implications for the theoretical design of 
materials with specific optical properties. 
}

%
Light propagating inside a heterogeneous solid experiences (i) absorption, and (ii) scattering.   
The light that is not absorbed is diffusely reflected and is responsible for the perceived color. 
The visual appearance of a material is hence determined by 
{\it  selective absorption of light} and {\it sufficient (back)scattering}.
For a material to be e.g. a luminous red pigment, two criteria must thus  
be satisfied. First, its absorption edge should be located 
at the appropriate energy ($\sim 2.1$~eV) so that the red component of the visible spectrum is not absorbed. 
Second, the absorption edge should be sharp, so that most other photons within the visible 
range (green, blue) are absorbed.

The computation of these effects from first principles  
is faced with three fundamental difficulties. 
First, in view of the sensitivity of the human eye, the optical gap must be obtained with a precision 
of at least $100$~meV. Conventional electronic structure methods yield a well-documented 
underestimation of the gap of conventional semiconductors. 
Second, the 
localized 4f states, which play a crucial role in optical properties of 
rare-earth based pigments\cite{hp-pigments-2002,Jansen2000,maestro1995}, 
are poorly described by standard density-functional theory\cite{dft} or even {\it GW} approaches\cite{ferdi_gw}. 
Third, a realistic assessment of the coloration of a pigment must take into account 
scattering properties depending on concentration, granularity and film thickness. 
{\it Ab initio} simulations so far have not ventured beyond
calculating the optical conductivity of infinite bulk samples
(see however Ref.~\cite{doi:10.1021/ja201733v,chromo_rubio} for organic molecules).
In this article we address all these issues and develop a general 
methodology for the prediction of the color of narrow-band materials. 

\begin{figure*}[!h!t]
\includegraphics[width=0.975\textwidth]{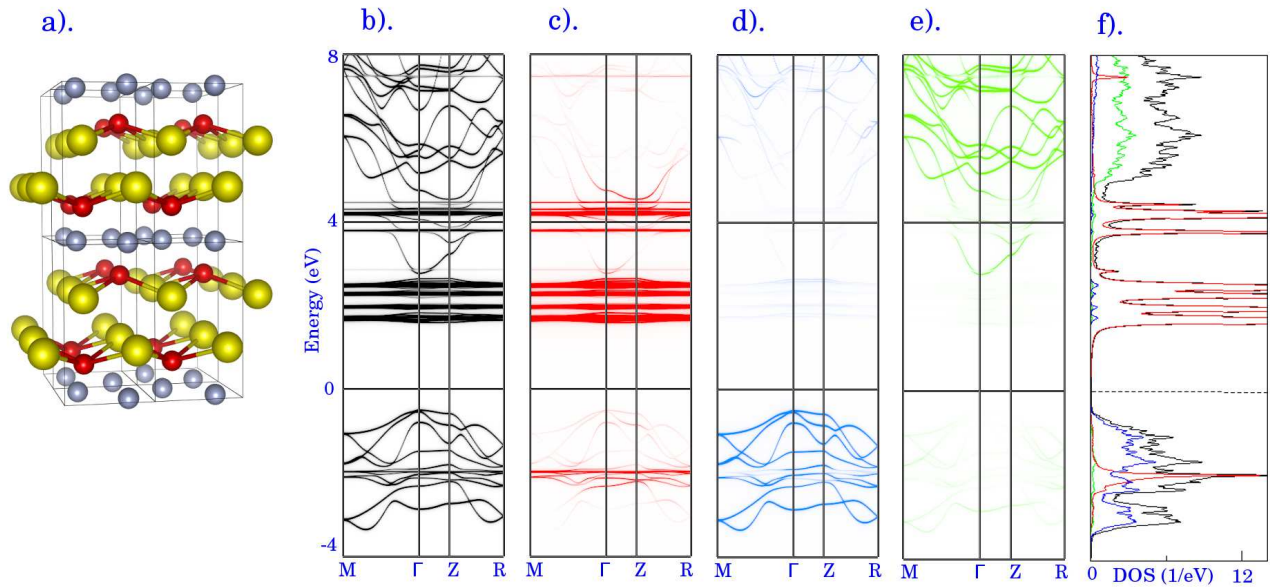}
\caption{{\bf Cerium fluorosulfide CeSF.} a) Crystal structure. The red, yellow and grey spheres represent Ce, S, and F, respectively.
b). The total many-body momentum-dependent spectral function $A_{\vk}(\omega)$ along high symmetry lines. 
c-e). The corresponding partial spectral functions for  Ce 4$f$ (c), S 3$p$(d), and Ce 5$d$(e).
f). The total momentum-integrated spectral function (black curve) as well as the partial momentum-integrated spectral functions for Ce 4$f$ (red),
S 3$p$ (blue) and Ce 5$d$ (green).}
\label{CeSF_bands}
\end{figure*}

%
We investigate cerium fluorosulfide (CeSF), a typical example of the new class 
of rare-earth pigments~\cite{Demourgues2001,Goubin2004}. 
It crystallizes in the layered ThCr$_2$Si$_2$ structure sketched in Fig.~\ref{CeSF_bands} 
(which is incidentally also that of the recently discovered iron-based superconductors~\cite{Kamihara2008}).
The same figure displays the momentum-dependent many-body spectral function 
$A_{\vk}(\omega)$ that encodes
the excitation energies associated with the addition or removal 
of an electron into the many-body ground-state (see Supporting Information).
The localized Ce-$4f$ states form quasi-atomic multiplets that hybridise weakly with the rest of  the solid. 
From the panels (c)--(e), displaying the specific orbital 
character of each electronic
state, it is apparent that the top of the valence band has 
dominantly S-$3p$ character.  
The occupied Ce-$4f$ states are located at 
higher binding energies, near the center of the S-$3p$ 
bands. The lowest unoccupied states are found to be the almost localized empty Ce-$4f$ states, 
with the dispersing Ce-$5d$ conduction band lying higher in energy.

\begin{figure}[!h!b]
  \begin{center}
\includegraphics[width=0.425\textwidth]{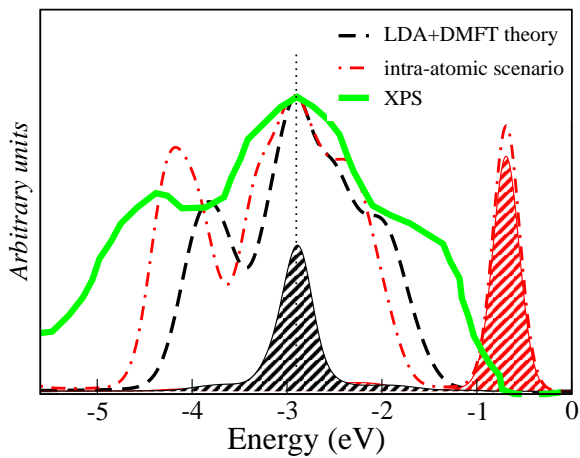}
    \caption{{\bf X-ray photoemission spectroscopy (XPS).} The many-body spectral momentum-integrated function (black curve) of CeSF compared to experimental XPS spectra \cite{pauwels_phd} 
(thick green curve).
In order to simulate experimental resolution the theoretical spectral function has been convoluted with a Gaussian of full width at half maximum of 0.3~eV.
As a comparison we show (red curve) the XPS spectrum of the alternative intra-atomic scenario previously proposed [See the text and Supporting Information for a discussion].
The shaded regions of the corresponding colors are the contribution from Ce 4f states in each case.
}
    \label{DOS_vs_XPS}
  \end{center}
\end{figure}

\begin{figure*}[h!t!]
  \begin{center}
\includegraphics[width=0.9\textwidth]{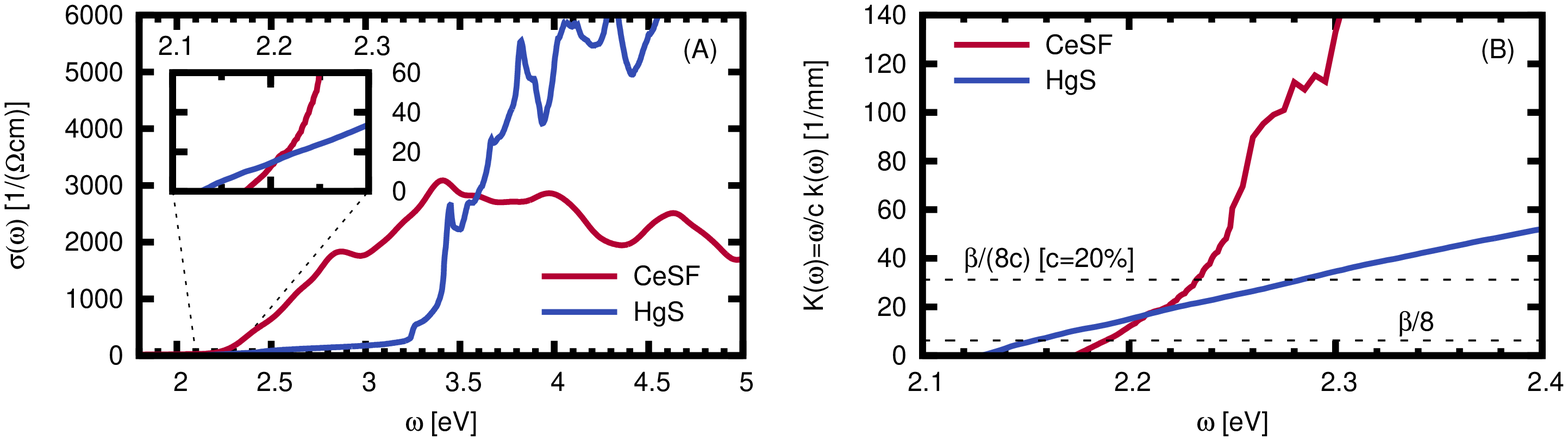}
  \end{center}
    \caption{{\bf Optical conductivity and absorption coefficient of CeSF and HgS.}
    (A) Polarization averaged optical conductivity. The inset is a zoom into the energy region of the onset of absorption.
(B) Macroscopic absorption coefficient $K$. The horizontal dashed lines are the quality criterion (\ref{criterion}) as defined in the text 
    computed for $\beta=50$/mm.
    }
    \label{figSK}
\end{figure*}

The calculated optical gap $\Delta=2.14$~eV ($578$~nm) is consistent with 
the red color of this material. Furthermore, our calculation reveals that the absorption 
edge is associated with the optical transition between S-$3p$ states and Ce-$4f$ states. 
This is at variance with the standard lore associating the coloration of cerium-based 
pigments to the intra-atomic Ce-$4f$ to Ce-$5d$ transition. 
Indeed, such a conventional assignment was proposed for CeSF in Ref.~\cite{Demourgues2001}, and 
X-ray photoemission spectra (XPS) were interpreted along the same lines in Ref.~\cite{pauwels_phd}. 
In \fref{DOS_vs_XPS}, we show that our results are nonetheless fully consistent with  
XPS measurements\cite{pauwels_phd}. We propose a different orbital assignment 
of the observed spectral peaks: according to our calculations,    
the peak at -3~eV  is due to contributions from localized Ce 4$f$ states 
while the shoulder at -1.5eV is mainly formed by S-3$p$ 
states that are pushed upwards in energy by the hybridization with Ce-4$f$ states.  
The previously proposed `intra-atomic' scenario can also be simulated theoretically 
(see Supporting Information) and is shown in Fig.~\ref{DOS_vs_XPS} to be inconsistent with the measured spectra.

In contrast to CeSF, mercury sulfide $\alpha$-HgS (also 
known as cinnabar red or ``vermilion'') has been used as a red pigment since antiquity \cite{vermilion-1993}.
It is a conventional semiconductor with a simple hexagonal structure. 
Although the additional complication of localized states is not present here, an accurate 
determination of the 
gap requires the use of {\it GW} calculations  (see Supporting Information 
for details). 
The onset of absorption is due to transitions between
broad and strongly hybridizing bands of mainly S-$3p$ and 
Hg-$6s$ character.
Given the qualitatively different nature of the optical transitions involved, 
HgS and CeSF are good pigments for entirely different reasons, 
as explained in detail below.

The absorption properties of these two compounds
can be derived from the frequency-dependent optical 
conductivity, $\mathrm{Re}\,\sigma(\omega)$. 
Using linear response theory, and neglecting excitonic effects (see the Supporting Information for a discussion), it can be expressed as
(see e.g.\ \cite{millis_review,optic_prb}):
\begin{eqnarray}\label{oc}
\mathrm{Re}\,\sigma_{\alpha} (\Omega)&=&\frac{2\pi e ^2}{V\hbar}\sum_{\svek{k}}
\int 
\dif\omega
\frac{\fermi{\omega}-\fermi{\omega+\Omega}}{\Omega} \nonumber \\
&&\times
\tr\biggl\{A_{\svek{k}}(\omega+\Omega)\hbar v_{\svek{k},\alpha}
A_{\svek{k}}(\omega)\hbar v_{\svek{k},\alpha} \biggr\}
\end{eqnarray}
where $V$ is the unit-cell volume, $\alpha$ labels the polarization, 
$v_{\svek{k},\alpha}$ are the optical transition matrix elements
and the Fermi functions $\fermi{\omega}$ restrict transitions 
to take place between occupied and empty states.
The absorption is described by the {\it macroscopic} absorption coefficient 
$K(\omega)$:
\begin{equation}
K(\omega)=\frac{\omega}{c_e}k(\omega)=\frac{\omega}{c_e}\mathrm{Im}\sqrt{1+\frac{4\pi i}{\omega}\sigma(\omega)}
\label{abs}
\end{equation}
where $k(\omega)$ is the imaginary part of the refractive index, and $c_e$ the speed of light.

\begin{figure*}[!h!t]
  \begin{center}
\includegraphics[angle=-90,width=0.5\textwidth]{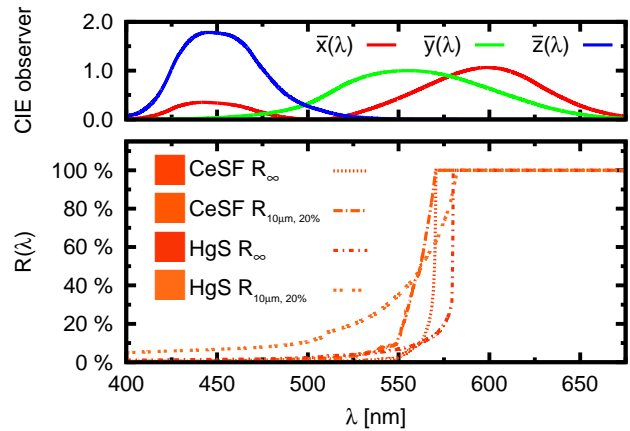}
    \caption{{\bf Diffuse reflectance and the colors} 
      Top: color-matching functions $\bar{x}$, $\bar{y}$, $\bar{z}$ of the human eye (CIE 1931 XYZ standard) as a function of wavelength $\lambda$  (see also Supporting Information).
    Bottom : calculated diffuse reflectances of CeSF and HgS. 
    Shown are the diffuse reflectances of the bulk ($R_{\infty}$)  and of a diluted (concentration $20\%$) thin film of 10$\mu$m on a white substrate ($R_{10\mu m,20\%}$) computed for  $\beta=$50~mm$^{-1}$.
    The corresponding color for each case is displayed in the legend).
}
    \label{figRl}
  \end{center}
\end{figure*}

The calculated optical conductivity and absorption coefficient of CeSF and $\alpha$-HgS are displayed in Fig.~\ref{figSK} (a),(b).
While marked differences exist between the two compounds on a broad energy range,    
the magnitude of $\sigma$ and $K$ near the onset of absorption (see the inset of \fref{figSK}(a) and \fref{figSK}(b)) is actually similar,
with $\sigma \simeq 50\,\Omega^{-1}\mathrm{cm}^{-1}$ and $K\simeq 20\mathrm{mm}^{-1}$ 
for $\omega=\Delta+0.1$~eV.
The physical origin of the spectral weight just above the absorption edge (which in turn determines $K$)  
is however fundamentally different for HgS and CeSF.  
Indeed we will now show that the large absorption in HgS is due to the strength of optical transition
matrix elements, while for CeSF it is due to the large density of  localized Ce-$4f$ as well as 
weakly dispersing S-$3p$ states.

To substantiate this claim, we note that color is determined by a fairly small frequency range 
above the absorption edge (see also below), and, hence, we focus on 
$\Delta \lesssim \hbar\omega \lesssim \Delta + \hbar\,\delomc$,
with $\hbar\,\delomc \approx 0.2\,\rm{eV}$.
In order to disentangle density of states effects from transition probabilities, we compute the ratio
between the optical conductivity, \eref{oc}, and the joint density of states $N(\omega)$~\footnote{obtained from 
replacing the trace in \eref{oc} with $\tr\left\{ A_{\svek{k}}(\omega+\Omega)\right\}\tr\left\{ A_{\svek{k}}(\omega)\right\}$, i.e.
by setting the Fermi velocities to unity and considering transitions between the total (traced) spectral weight.%
} 
at the energy $\omega=\Delta+\hbar\,\delomc$. 
We find that $\sigma/N=2.5\, (\hbar/(a_0m_e))^2$ for HgS, while $\sigma/N=0.86\, (\hbar/(a_0m_e))^2$ 
is almost three times smaller for CeSF ($a_0$ is the Bohr radius and $m_e$ the mass of the electron).
This proves that transition matrix elements dominate for HgS, while 
density of states effects dominate for CeSF. 
The reason for this is the weak dispersion of the Ce-$4f$ states (expected from their localized character), 
but also of the S-$3p$ states at the top of the valence band in CeSF. 
Interestingly, the weak dispersion of the S-$3p$ states along the $c$-axis is due to the quasi two-dimensional
environment of the sulfur atoms, which are located within layers parallel to those containing cerium atoms 
(Fig.~\ref{CeSF_bands}(a)). 
As detailed in the Supporting Information, the joint density of states $N(\omega)$ and the absorption $K(\omega)$ display
a discontinuous increase at threshold $N\propto\theta(\hbar\omega-\Delta)$ for a strictly two-dimensional dispersion. 
In contrast, a three-dimensional parabolic dispersion (mimicking e.g. the 3p or 6s bands in HgS) 
yields a less sharp frequency-dependence $\propto(\hbar\omega-\Delta)^{1/2}$ above the absorption edge. 

The macroscopic quantities $\sigma$ and $K$ describe the absorption properties of a 
perfectly crystalline bulk solid. 
Pigments are used however (in e.g. paints) in the form of small particles embedded
into a transparent matrix.
In this context, light propagation depends on the morphology of the sample, 
and a multiple scattering problem has to be solved to determine the {\it diffuse reflectance}  $R(\omega)$ 
(in contrast to the simpler specular reflectivity~\cite{me_psik,optic_epl}). 
A commonly used~\cite{Levinson2005_2} 
approach is the effective-medium description 
due to Kubelka and Munk (KM)~\cite{Kubelka1948,Levinson2005}.
The KM model treats the propagation of light through a homogeneous layer of 
matter with pigment concentration $c$, that absorbs light with rate $cK(\omega)$ 
and backscatters with rate $\beta(\omega)$ per unit length. 
The 
quantity $\beta$ effectively contains all the information on the microscopic structure of the sample. 
As shown in the Supporting Information, the energy dependence 
of $\beta$ can be neglected, and we 
use a typical
value $\beta = 50\rm{mm}^{-1}$, consistent with
the range of values $\beta \approx 10-500 \rm{mm}^{-1}$
reported in the literature for a wide range of industrial inorganic pigments~\cite{Levinson2005_2}. 
For a semi-infinite layer, the KM model yields 
a simple expression for the diffuse reflectance: 
$R_\infty(\omega)=\alpha(\omega)-\sqrt{\alpha(\omega)^2-1}$, with
$\alpha(\omega)=1+2cK(\omega)/\beta(\omega)$.
As expected, $R_\infty\approx 1$ for $cK/\beta\ll 1$ (weak absorption) and
$R_\infty\approx 0$ for $cK/\beta \gg 1$ (strong absorption). The formula for a sample of finite thickness is given in
the Supporting Information.

We have employed the KM model in conjunction with our first-principles absorption $K(\omega)$ 
to compute the diffuse reflectance of CeSF and HgS samples, as well as their 
color (Fig.~\ref{figRl}). 
The latter is obtained (e.g. as $XYZ$ tristimulus values or $xy$ chromaticities) by taking into account the spectral 
distribution of the light source as well as the sensitivity of the human eye to red, green and blue light as 
encoded in the empirical color matching functions displayed in the top panel of Fig.~\ref{figRl} (for details, 
see Supporting Information). 
We considered semi-infinite samples as well as realistic $10\mu$m layers on a white ($R$=1) substrate, 
consisting of either pure ($c=100\%$) or diluted  ($c=20\%$) pigments.
The resulting diffuse reflectances and colors depicted in Fig.~\ref{figRl} reveal that the pure semi-infinite
bulk samples of both materials have bright red colors, while the thin films with a pigment concentration of $20\%$
have a more orange tone, especially for HgS.

We now introduce a simple performance criterion for the usability 
of such materials as pigments. 
In order for the color to be a bright red, the reflectance for $\omega>\Delta$ should drop 
sufficiently quickly such that admixtures from the green and blue part of the visible spectrum 
are suppressed. This can be insured by requiring that $R(\Delta+\hbar\delta\omega_c){\le}1/2$. 
Here, $\hbar\delta\omega_c\approx 0.2\rm{eV}$
is the characteristic frequency interval over which the human eye distinguishes 
between the primary colors (corresponding to $\delta\lambda\approx 60$nm for red light, see 
top panel of Fig.~\ref{figRl}).
Since the reflection for $\omega>\Delta$ arises from a finite absorption,
the above requirement can be translated, using the KM model, into the criterion:
\begin{equation}
K(\Delta+\hbar\delta\omega_c)\stackrel{!}{\ge}\beta/(8c)
\label{criterion}
\end{equation}  
This threshold is marked in \fref{figSK}(b): for pure materials ($c=100\%$)
both CeSF and HgS largely satisfy the criterion. Indeed, $R_\infty(\lambda)$ with $\beta=50\rm{mm}^{-1}$ switches from
$R_\infty=0$ to $R_\infty=1$ within only a few nanometers (Fig.~\ref{figRl}).  
Criterion (\ref{criterion}) can be turned, for a given $c/\beta$, into a minimal thickness  
$L_{min}$ above which the pigment has the same apparent color as the bulk, i.e.\
its color is sufficiently stable. This is  represented in Fig.~\ref{km_univers}.  
Using $\beta=50/mm$, the minimal thickness of a paint layer consisting of $20\%$ CeSF is 3.2$\mu$m,
whereas it is 35$\mu$m for $20\%$ HgS.
As a result, the reflectivity of $10\mu$m layers with 20\% diluted HgS violates 
our quality criterion, which explains the notably different (orange) color of those samples, observed in \fref{figRl}.

\begin{figure}[!t]
  \begin{center}
\includegraphics[angle=-90,width=0.45\textwidth]{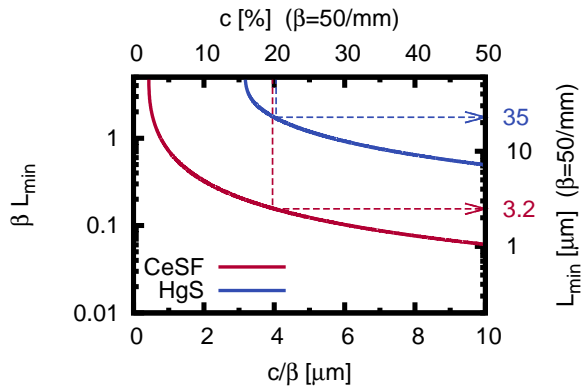}
    \caption{{\bf Universal quality curves for thin films of CeSF and HgS.} The thin films located above the corresponding curve in the ($\beta L_{min}$,$c/\beta$) coordinates pass the quality
criterion for a good pigment (see the text). 
    The top and right axes correspond to a fixed value of $\beta$ of $50$/mm, allowing for a direct translation of a given concentration into a minimal film thickness.
    }
    \label{km_univers}
  \end{center}
\end{figure}

In conclusion, we presented a framework for the theoretical determination of the color of pigment materials. 
Our methodology combines first-principles calculations of the frequency-dependent absorption based on 
state-of-the art many-body theories, with an intermediate length-scale modelization of the scattering properties of realistic 
samples. 
We applied this methodology to classic `vermillion' HgS and to the recently discovered rare-earth pigment CeSF. 
Our results reveal that the red coloration of these two materials has a very different origin at the atomic scale. 
While it is due to large optical matrix elements in HgS, the key effect in CeSF is the large 
density of weakly dispersing states (resulting both from the low-dimensionality of the crystal 
structure and the localized character of Ce $4f$ states). 
We also demonstrated that the relevant optical transitions in CeSF are S-$3p$ to Ce-$4f$ inter-atomic transitions, 
in contrast to previous belief. 
Our methodology may lead to rational materials design in which first-principles calculations are used 
in synergy with solid-state chemistry in order to create 
new
materials with specific optical properties.

\begin{small}

\section{Methods}

Combining electronic  
structure and many-body techniques \cite{geo96,ani97,lichtenstein_lda+dmft_1998,SCDMFT,Aichhorn2009,Aichhorn2011,triqs} we calculate the electronic structure of rare-earth pigments. 
Then we take these approaches further by performing an 
{\it ab initio} calculation
of the optical response functions and going all the way
to a prediction of the actual color 
(expressed e.g. in RGB coordinates \cite{cie,1475-4878-33-3-301})
of a sample with a given granularity, using an intermediate 
length-scale modelisation \cite{Kubelka1948}.

As discussed in the main text,
the strong Coulomb repulsion between the $4f$ electrons in CeSF
is responsible for
the splitting between the empty Ce-$4f$ states at the bottom of 
the conduction band and the occupied Ce-$4f$ states,
which hybridise with the S-$3p$ states
to form the valence band.
A proper treatment of this strong Coulomb
interaction is essential in order to correctly reproduce the optical
gap and electronic structure of this material.
Hence, we have employed 
the so-called LDA+DMFT approach\cite{ani97,lichtenstein_lda+dmft_1998}, which combines
electronic structure calculations with a many-body treatment of local correlations on the Ce 4$f$ shell
in the framework of dynamical mean-field theory (DMFT) \cite{geo96}.
The interaction vertex has been computed from first principles using the constraint random phase approximation\cite{PhysRevB.70.195104,Loig2012}.

Within LDA+DMFT, the Ce-$4f$ shell is treated as that of an effective atom,
self-consistently coupled to an environment describing the rest of the solid.
From the Hamiltonian of this effective atom (which takes into account crystal-field
effects, intra-atomic Coulomb interactions and the spin-orbit coupling),
a many-body self-energy is computed and inserted
into the Green's function of the full solid. Self-consistency over
the total charge density and the effective atom parameters is implemented~\cite{Aichhorn2009,Aichhorn2011}.
Important technical points of our calculational approach are presented
in the Supporting Information.

\begin{acknowledgments}
We are grateful to Erich Wimmer and Ren\'e Windiks, who
introduced us to the field of rare-earth-based pigments, and acknowledge 
useful discussions with David Jacobs and Hong Jiang.
This work was supported by the French ANR under projects
CorrelMat and SURMOTT, by the NSF materials world network under
Grant NSF DMR 0806937,
as well as by IDRIS/GENCI under
project 1393.
\end{acknowledgments}

\paragraph{author contributions}
J.~M.~T. and L.~P. performed 
the calculations of the electronic structure and optical spectra. 
L.~V. performed the calculations of the screened Coulomb interactions.   
S.~B. and A.~G. provided guidance and coordinated the project. 
All authors contributed to the analysis and discussion of the results, and to the writing of the article. 

\end{small}


\newpage

\noindent
{\bf 
Rare-Earth vs. Heavy Metal Pigments\\ and their Colors from First Principles --\\ Supporting Information
}\\
Jan M. Tomczak\\
L. V. Pourovskii\\
L. Vaugier\\
A. Georges\\
S. Biermann\\

\medskip

\maketitle

{\bf 
These supporting notes give additional information concerning
the calculation of color of CeSF and HgS. In particular, we 
(a) describe the technical details of our computational scheme, 
``LDA+DMFT'', the combination of density functional theory within 
the local density approximation (LDA) with dynamical mean field 
theory (DMFT),
(b) present the momentum resolved electronic structure and optical
conductivity of our target compounds CeSF and HgS,
(c) provide further evidence for our interpretation of the onset
of optical transitions in CeSF being determined by {\it p-f} transitions
by discussing the implications on the electronic structure and
optical properties that would result from an alternative ``{\it f-d} scenario'',
(d) summarize the Kubelka-Munk theory, which allows us to 
deduce the color of our compounds from the knowledge of the
bulk optical conductivity, including practical details of our 
intermediate length scale modelisation.
We conclude with a discussion of the {\it shape} of the onset
of the absorption spectrum for the two prototypical compounds
that we investigate, the quasi-two-dimensional CeSF with its
localized f-electrons and the three-dimensional, more conventional
HgS with broad and well-hybridising bands.
}

\section{Combined density functional and dynamical mean-field 
simulations of CeSF}

Our combined density functional and dynamical mean field ``LDA+DMFT''
implementation \cite{Aichhorn2009,triqs} is based on a full-potential 
linearized augmented plane waves (LAPW) technique
as implemented in the Wien2k code \cite{Wien2k}. 
Our calculations are fully self-consistent, including over the
charge density \cite{Aichhorn2011}. 
Spin-orbit coupling has been taken into account through a
second variational treatment. 

The LDA+DMFT calculations for CeSF have been carried out for 
the paramagnetic phase at the experimental crystal structure
\cite{Demourgues2001}; it is of tetragonal symmetry (space group
$P4/nmm$), with the lattice parameters $a$=3.993\AA\ and $c$=6.950\AA.
The correlated $f$ orbitals are constructed from projected atomic 
orbitals which are promoted to Wannier function through an 
orthonormalisation procedure. In practice, we have
used an energy window $[-8:10.8]$ eV, which includes bands with 
mainly $f$ orbital character, following the method implemented for 
$d$ electron materials in Ref. \cite{Aichhorn2009}. 
Local correlations between the Ce 4$f$ states have been treated within DMFT. 
The Hubbard and Hund interactions, $U=4.8$ eV and $J=0.70$ eV, 
resulting from the local Coulomb interaction on the Ce 4$f$ shell, 
have been calculated within the constrained Random Phase Approximation 
(cRPA) \cite{PhysRevB.70.195104}. 
We have employed the recent implementation of cRPA into the LAPW framework 
of the Wien2k code of Refs.~\cite{Loig2012,Loig-thesis}. 

\begin{figure*}[!t]
\begin{center}
\includegraphics[width=12 cm]{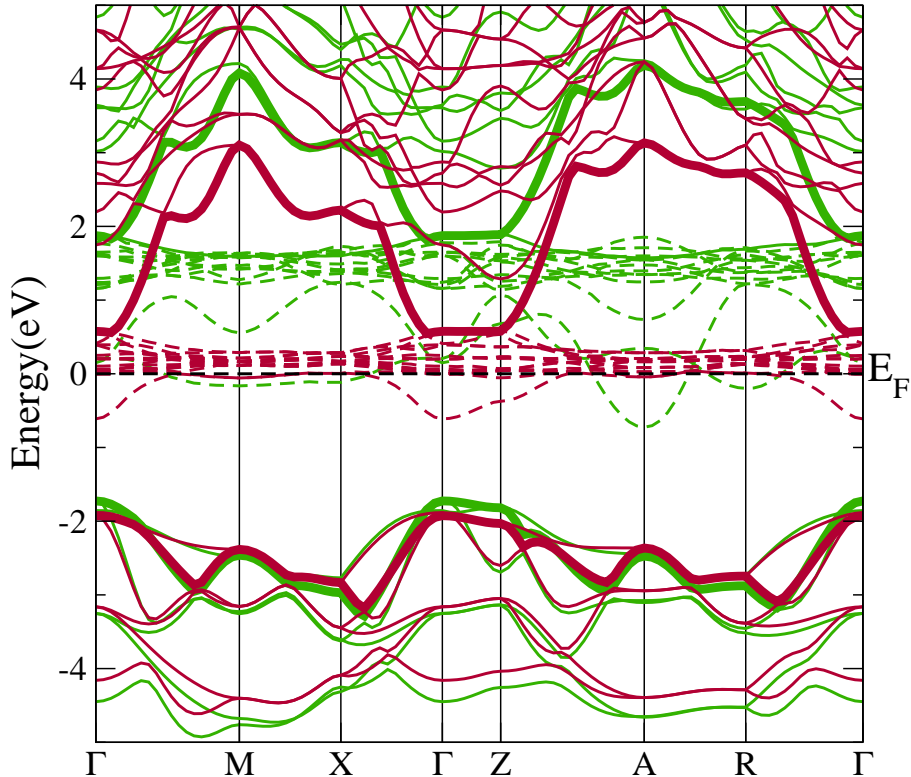}
\caption{{\bf The band structure of CeSF in LDA and {\it GW}.} Band-structure of CeSF obtained within LDA (in red) and {\it GW} (in green). The Ce 4$f$ states were treated as valence states in both cases.
The bands of Ce 4$f$ character are shown by dashed lines. The topmost S 3$p$ and lowest Ce 5$d$ bands are highlighted by fat lines. One may clearly see that the {\it GW} correction affects the Ce5$d$ bands much more strongly then S3$p$ ones. 
}\label{fig:GW_LDA_CeSF}
\end{center}
\end{figure*}

The Hubbard-I approximation has been employed as a quantum impurity
solver within the LDA+DMFT scheme. We have used the full four-index Coulomb 
interaction matrix in the Hubbard-I impurity solver, thus all multiplet 
effects within the Ce 4$f$ shell were taken  into account. 
Double counting (DC) of Coulomb interactions between the $f$ electrons
was avoided, using the fully-localized limit (FLL) expression  
$\Sigma_{\sigma}=U(N-1/2)-J(N/2-1/2)$, where $N \approx 1$ is the 
occupancy of the Ce 4$f$ shell in the DMFT quantum impurity 
problem\cite{SCDMFT}. 
In the self-consistent LDA+DMFT calculations we employed 225 
${\bf k}$-points in the irreducible part of the tetragonal Brillouin 
zone (BZ) for the BZ-integration. All calculations have been carried
out at the temperature of 116~K with a grid of 1024 Matsubara points, 
a high-frequency expansion up to the tenth order in 
$1/(i\omega_n)$
has been employed to describe tails of the Green's function and 
the self-energy at Matsubara frequencies above the cutoff.

\begin{figure*}[t!]
\begin{center}
\includegraphics[width=14.2 cm]{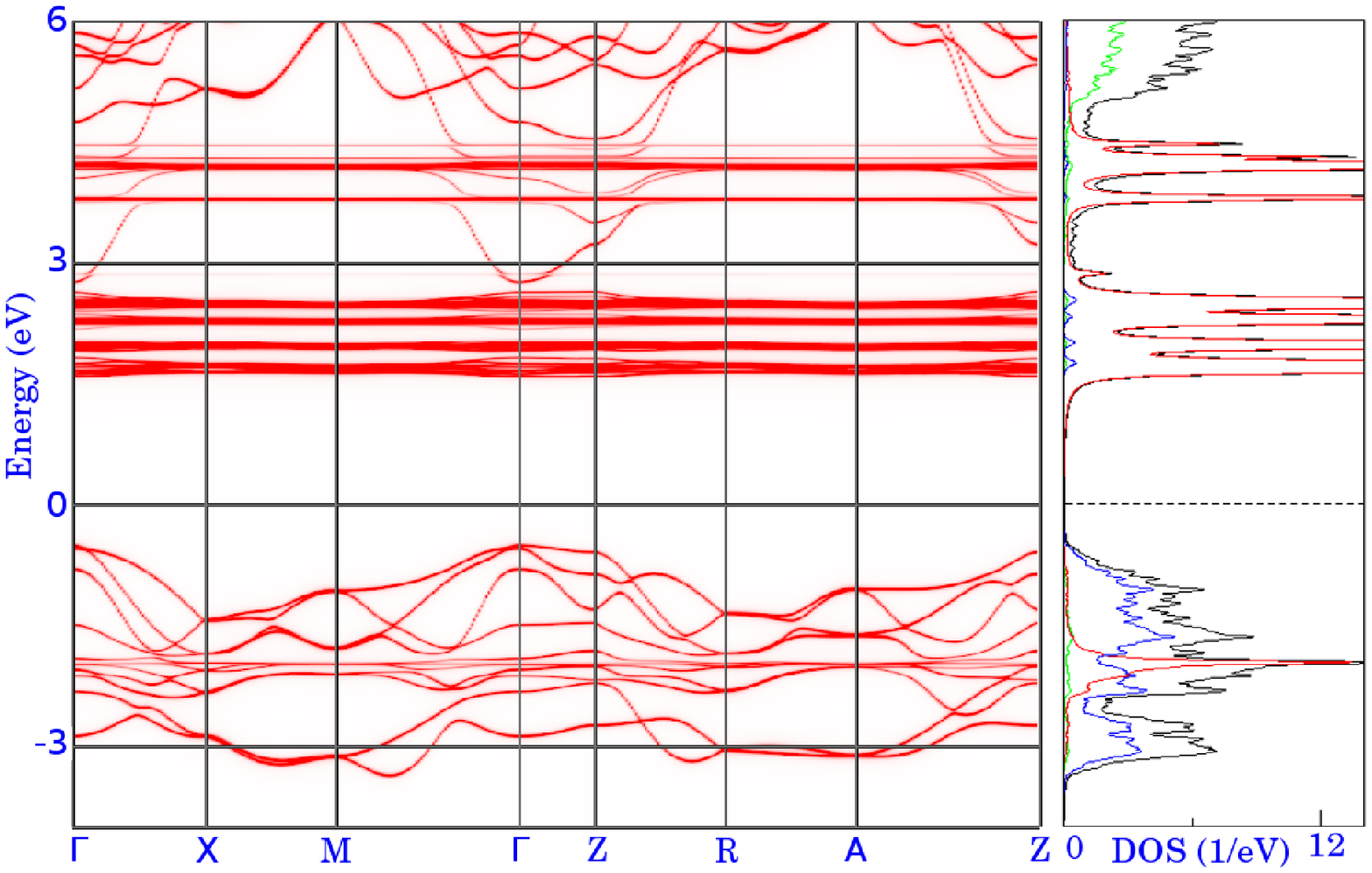}
\caption{{\bf The LDA+DMFT excitation spectrum.} k-resolved spectral function (left panel) and the total (black), partial Ce 4$f$ (red),
 Ce 5$d$ (green), S 3$p$(blue) density of states (right panel) for CeSF}\label{CeSF_bands_full}
\end{center}
\end{figure*}

The local density approximation leads to a rather significant, yet 
expected, underestimation of the semiconducting gap between S 3$p$ 
and lanthanide 5$d$ bands in the rare-earth fluorosulfides. 
The LDA+DMFT approach employed in this work does not contain corrections 
for this LDA error, which is due to long-range interactions and
the missing derivative discontinuity in the Kohn-Sham band structure.
In addition, CeSF contains a partially filled localized 4$f$ shell, 
and as pointed out in the main text,
the relative positions of S 3$p$, Ce 5$d$ and Ce 4$f$ states define
the color of the material. In order to evaluate the effect of long-range 
interactions on the relative position of S 3$p$ and Ce 5$d$ bands we 
have performed one-shot calculations within Hedin's {\it GW} approximation\cite{ferdi_gw,RevModPhys.74.601}
-- using the implementation described in detail in Refs.~\cite{PhysRevLett.102.126403,PhysRevB.82.045108} --
for CeSF and, for sake of comparison, for LaSF. For the latter, the 
optical gap is known to open between the valence S 3$p$ and conduction 
La 5$d$ bands. The band gap in LaSF calculated within the LDA is equal 
to 1.3~eV compared to the experimental optical gap of 2.8~eV 
\cite{Demourgues2001}. 
Within the {\it GW} approximation, the value of the gap is 
increased by about 1.17 eV compared to the LDA one, and, hence, the 
LDA error is compensated by about 80\%. As in usual semiconductors,
the shape of
the highest valence and lowest conduction bands, which are crucial for 
the absorption edge, are barely affected by {\it GW} corrections.

\begin{figure*}[t!]
  \begin{center}
\includegraphics[width=16cm]{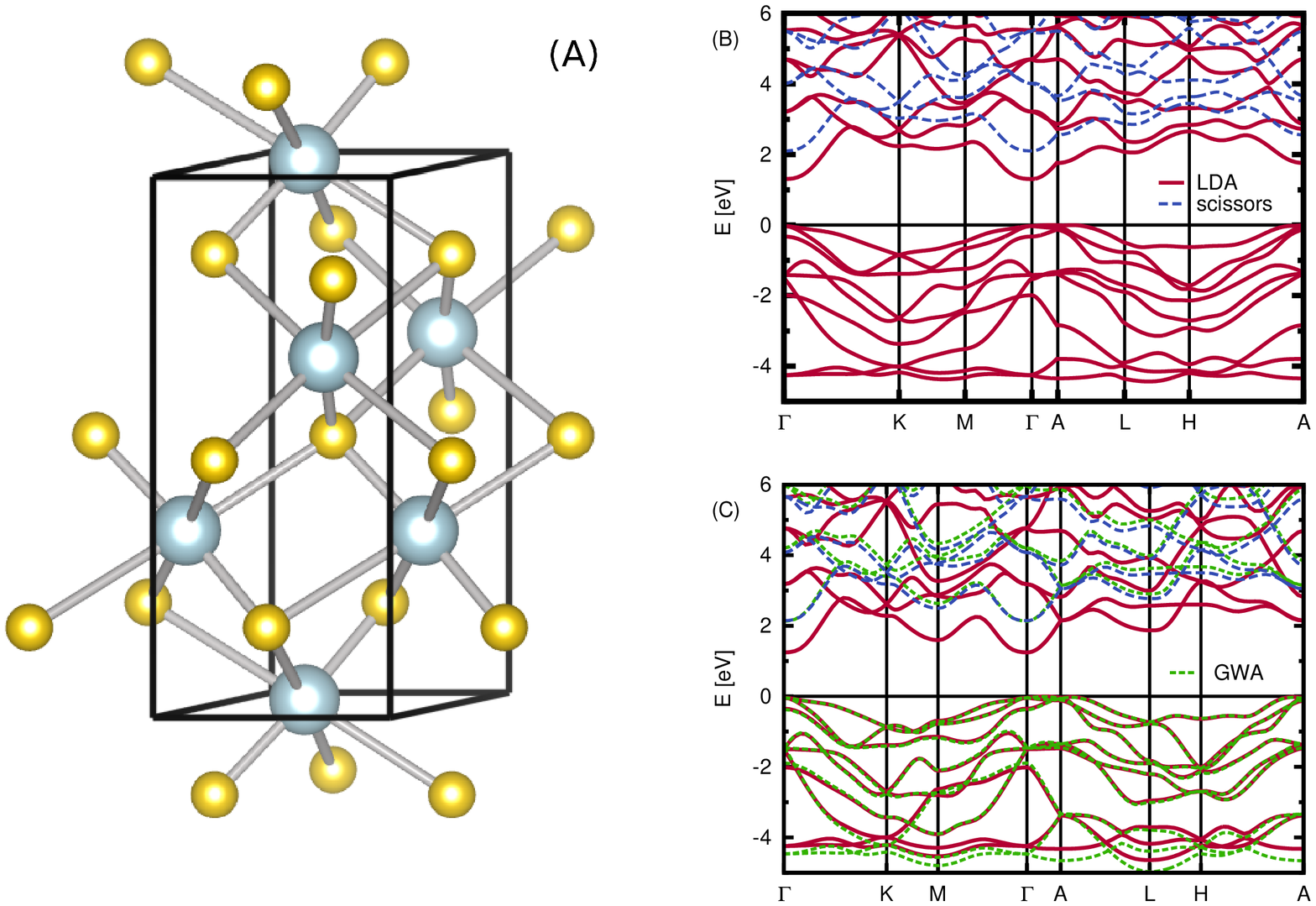}
      \caption{{\bf Crystal and band-structure of $\alpha$--HgS.} (A) crystal structure of $\alpha$-HgS (mercury atoms in blue, sulfur in yellow,
      graphic produced with Ref~\cite{vesta}). (B)-(C) Shown are results of (i) DFT-LDA using Wien2k (red, panel (B)) and FPLMTO (red, panel (C)), (ii) the {\it GW} approximation (green, panel (B)),
      and (iii) rigidly shifted (``scissors operator'') LDA conductions bands (blue, panel (B)~:  Wien2k, panel (C)~: FPLMTO).}
      \label{hgsbnds}
      \end{center}
\end{figure*}

We have performed similar {\it GW} simulations\cite{PhysRevLett.102.126403,PhysRevB.82.045108} for CeSF, and the resulting  
{\it GW} $\vk$-resolved spectral function is compared to the LDA band structure 
in Fig.~\ref{fig:GW_LDA_CeSF}.
As expected, the {\it GW} method is not able to describe correctly the 
strongly-correlated Ce 4$f$ states, which remain metallic with their 
bandwidth increased relative 
to the LDA one. The important thing to notice here is that the upward shift for the bands of Ce 5$d$ character 
relative to $E_F$ is approximately 6 times larger then the corresponding 
(mostly downward) shift of the S 3$p$ bands. Therefore,
one may conclude that the underestimation of the pd-gap in this
case is mainly an effect of the missing derivative discontinuity
in the Kohn-Sham description, and thus a too low position of 
the 5$d$ conduction states. As in the case of LaSF, the shape 
of the topmost valence and lowest conduction bands is weakly affected 
by {\it GW} corrections, and can be well described as a rigid shift of the 
corresponding bands. The value of the gap between those states is 
increased by of about 1.15 eV, similarly to the case of LaSF.

On the basis of our {\it GW} simulations we conclude that, in the case of
CeSF, corrections to the Kohn-Sham band structure of the ``uncorrelated''
3p- and 5d-bands 
and thus the relative positions of the CeSF bands relevant for 
its absorption edge -- can be well described
by a rigid shift of the Ce 5$d$ states relative to Ce 4$f$ and S 3$p$ 
by a value close to the LDA band gap underestimation in LaSF.
Hence, we have applied a static shift of 1.5 eV to the Kohn-Sham 
eigenvalues for unoccupied states with respect to the occupied bands 
and partially occupied Ce 4$f$ states in our LDA+DMFT simulations of CeSF. 

\section{The LDA+DMFT k-resolved spectral function for CeSF}

The $\vk$-resolved spectral function $A_{\vk}(\omega)$
encodes the excitation energies associated with the addition or removal of an electron on top of the many-body ground-state $\Phi_0$:

\begin{eqnarray}
A_{\vk}(\omega)&=&\sum_{\Gamma} \left\langle \Phi_{\Gamma}\left|c^{\phantom{\dagger}}_\vk\right|\Phi_0\right\rangle^2 \delta(E_{\Gamma}-E_0+\omega+\mu)\nonumber \\
&+& \sum_{\Gamma} \left\langle \Phi_{\Gamma}\left|c^{\dagger}_\vk\right|\Phi_0\right\rangle^2 \delta(E_0-E_{\Gamma}+\omega+\mu),
\end{eqnarray}
where 
the sum runs over all excited many-body states $\Phi_{\Gamma}$ with energies $E_{\Gamma}$, $\mu$ is the chemical potential. When correlations are absent the many-body states $\Phi_0$ and $\Phi_{\Gamma}$ are just Slater determinants, and the corresponding excitation energies are one-electron energies of the corresponding Hartree-Fock (or Kohn-Sham) problem.

\noindent
In the framework of LDA+DMFT and in the basis of Kohn-Sham eigenstates the spectral function $A_{\vk}(\omega)$ reads:
\beq
A_{\vk}(\omega)=-\frac{1}{\pi}Im \left[Tr \left( \omega+i\eta+\mu-\hat{\epsilon}_{\vk}-\Sigma_{\vk}(\omega) \right)^{-1} \right],
\eeq
where $\hat{\epsilon}_{\vk}$ is the diagonal matrix of Kohn-Sham eigenstates, $\Sigma_{\vk}(\omega)$ is the DMFT local self-energy upfolded
into the basis of Kohn-Sham eigenstates \cite{Aichhorn2009} (due to this upfolding it acquires $\vk$-dependence) and $\eta$ is a positive infinitesimal. 
When correlations beyond DFT are not included ($\Sigma=0$) the $\vk$-resolved spectral function has $\delta$ peaks at Kohn-Sham eigenvalues that form
one-electron Kohn-Sham bands. In the case of metallic systems with weak or moderate correlations the Kohn-Sham eigenvalues are shifted and the corresponding $\delta$ peaks broadened
due to quasi-particle renormalization and life-time effects, respectively. In the present case of strongly-correlated Ce 4$f$ states the DMFT self-energy leads to the destruction
of coherent metallic 4$f$ states and the formation of 
lower and upper Hubbard bands, hence opening the semiconducting gap.

The $\vk$-resolved spectral function of CeSF along the $M-\Gamma-Z-R$ path in the Brillouin zone is shown in Fig.~1 of the main text.
In Fig~\ref{CeSF_bands_full} we show the LDA+DMFT spectral function for CeSF for a longer ($\Gamma-X-M-\Gamma-Z-R-A-M$) path along
with the corresponding integrated spectral function.

\section{Band-structure calculation of HgS}

For cinnabar red, $\alpha$--HgS, we calculate, as a first step, the Kohn-Sham 
band-structure within the local density approximation (LDA), using the Wien2k\cite{Wien2k} package.
The crystal structure of $\alpha$--HgS is hexagonal, with space group P$3_221$, 3 formula units per cell, and lattice constants a=4.145\AA\ and c=9.496\AA\cite{hgs_xstruc}
\footnote{In our calculation we neglect the spin-orbit coupling, since its influence at low energy was found to be small\cite{PhysRevB.82.085210}.}.

The resulting band-structure is shown in \fref{hgsbnds} (a). 
We find an indirect gap of 1.3eV (from valence bands near the A-point to the conduction band at $\Gamma$), and an optical (=direct) gap of 1.33eV at the $\Gamma$--point.
These findings are consistent with earlier results using the generalized gradient approximation.\cite{sun:113201}
Owing to the well-established underestimation
of insulating gaps within density functional theory, these values are much smaller than the experimental optical gap of $\sim2.1$~eV\cite{hgs_choe}.

In these cases it has become customary (see e.g. the recent Ref. \cite{PhysRevB.82.085210}) to employ a scissors operator to tune the gap to its experimental value,
before proceeding with the determination of e.g.\ optical properties.
However, this procedure is {\it ad hoc} and would deprive us of our claim of performing {\it ab initio} calculations.
Therefore, we went a step further and applied the {\it GW} approximation\cite{hedin} to HgS. {\it GW}, a true many-body approach, has proved to be a valuable tool for moderately correlated systems, and is particularly suited to correct band-gaps in semi-conductors\cite{ferdi_gw,RevModPhys.74.601} (See also our paragraph on {\it GW} applied to CeSF in these supporting information).

\begin{figure*}[!t!h]
  \begin{center}
   {\includegraphics[angle=-90,width=.6\textwidth]{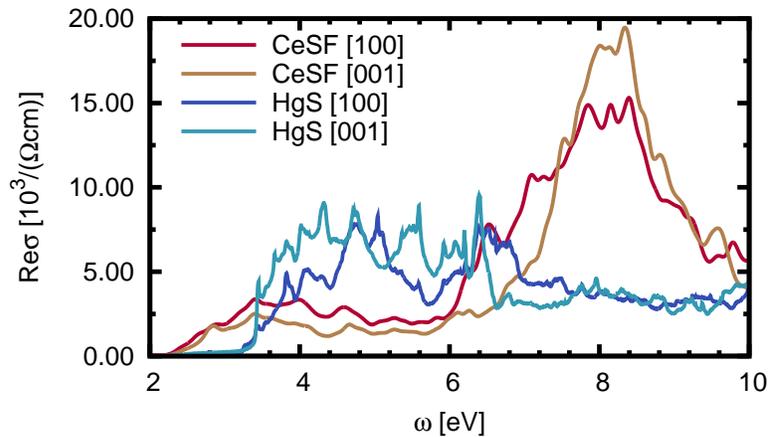}}
\caption{{\bf Optical conductivities of CeSF and HgS.} Shown are data over a wider energy range, and for different light polarizations with respect to the crystallographic axes.}
      \label{optcond2}
      \end{center}
\end{figure*}

For technical reasons, we resorted to a different electronic structure technique, the full potential linear muffin tin orbital (FPLMTO) method\cite{fplmto}.
Using again the LDA potential, the FPLMTO band-structure, shown (in red) in \fref{hgsbnds} (b) is in reasonable agreement with our results using Wien2k. 
Taking these FPLMTO findings as starting point, we performed a one-shot {\it GW} calculation using the code of Ref.\cite{fpgw}.
 
The resulting quasi-particle band-structure [as obtained in the usual way of perturbing the Kohn-Sham eigenvalues to linear order in frequency in the {\it GW} self-energy\cite{ferdi_gw}]
is shown (as the green dashed lines) in \fref{hgsbnds} (b).
The indirect gap comes out to be 2.1eV, in excellent agreement with experiment\cite{hgs_choe}.
In the occupied part, this {\it GW} band-structure closely follows the LDA result for energies larger than -2eV.
In the figure are also shown (in blue) the LDA conduction bands rigidly shifted upwards by a constant value. Evidently, with this scissors operation 
also the conduction bands of the LDA yield a favorable approximation to the {\it GW} results over at least 1eV of the bandwidth of unoccupied states.
Therewith the ``scissored'' LDA covers an energy range sufficient to account for the onset of optical absorption.

With this {\it ab initio} justification, we turn back to the Wien2k data and apply a scissors operator
to reproduce the {\it GW} gap, yielding an optical gap of 2.13eV. The resulting shifted conduction bands are shown (again in blue) in \fref{hgsbnds} (a).
This establishes the band-structure that our calculations of
the optical conductivity of HgS are based on.

It is worth mentioning that the onset of optical absorption
takes place between bands of mainly S3p character for the valence
band and Hg6s with admixtures of Hg6p and S3p for the conduction
band.

\section{Optical conductivities}

In the main text, we showed the low energy onset of absorption 
in the polarization averaged optical conductivity. Owing 
to the non-cubic crystal structures (CeSF: tetragonal, HgS: hexagonal) 
there is a dependence of the optical response on the orientation of the light polarization with 
respect to the crystallographic axes. \Fref{optcond2} displays our results 
resolved into the two inequivalent orientations
(along the crystallographic a and c-axis), as well as for an expanded energy range. 
In both compounds, the polarization dependence is notable mostly at higher energies.
In the case of $\alpha$--HgS we find an overall agreement with the dielectric function of Ref. \cite{PhysRevB.82.085210}\\

\medskip

We further would like to comment on our assumption that excitonic effects are negligible in the compounds considered,
thus verifying our methodology (For a general discussion of these effects see Ref.~\cite{RevModPhys.74.601}).

In conventional semiconductors, excitonic effects are well-known to severely modify absorption spectra
through the appearance of sharply defined bound states below the conduction bands
(see e.g.\ Ref.~\cite{PhysRevLett.85.2613} for the example of SiO$_2$). 
Such features are absent in the experimental dielectric function of both HgS\cite{PhysRevB.82.085210}, and CeSF\cite{Goubin2004}.
Besides, the electron-hole attraction in the excitonic regime may red-shift the onset of absorption
(See e.g.\ Ref.~\cite{PhysRevLett.81.2312} for a discussion on GaAs). 
Since we obtain good values for the charge gaps for both HgS and CeSF, however, we believe this effect to be minor.
Of course there can be effects of cancellation, e.g.\ in the case of HgS, between the electron-hole interaction induced redshift, 
and a potential gap {\it over-estimation} within {\it GW}. However, experience shows that LDA based one-shot {\it GW} calculations
correct well for underestimations of LDA band-gaps, and rarely yield a too large gap with respect to experiment\cite{fpgw}.

\section{{\it f-d} intra-atomic scenario for the absorption edge in CeSF: electronic structure}

\begin{figure*}[!t]
\begin{center}
\includegraphics[width=14 cm]{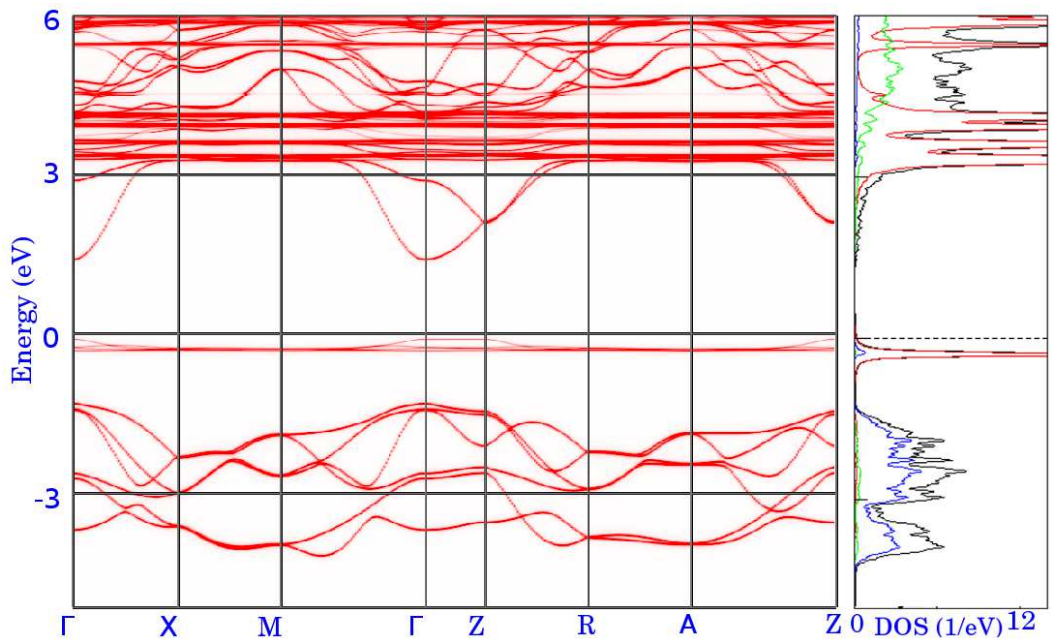}
\caption{{\bf The {\it f-d} scenario.} The LDA+DMFT band structure of CeSF (left panel) obtained in the {\it f-d} scenario. Right panel: the corresponding total (black), partial Ce 4$f$ (red),
 Ce 5$d$ (green), S 3$p$(blue) density of states }\label{pshift1}
\end{center}
\end{figure*}

Similarly to many other Ce-based semiconductors (for example, the sesquioxide Ce$_2$O$_3$ \cite{Golub95}, or the sesquisulfide Ce$_2$S$_3$ \cite{Dagys1995,Witz1996, Windiks2008438}) 
the absorption edge in CeSF has been previously ascribed to the Ce intra-atomic 4$f$ to 5$d$ transitions\cite{Demourgues2001}.
We will call this the ``{\it f-d} scenario'', in contrast to the true ``{\it p-f}'' scenario established by our {\it GW} calculations for CeSF (see above).
 As discussed in the present manuscript, our LDA+DMFT
calculations lead thus to a rather different picture of the CeSF electronic structure, with the onset of absorption being
due to optical transitions between the occupied S 3$p$  and empty Ce 4$f$ bands, which result in a sharp absorption
edge and the bright red color of this compound.

However, we have also simulated the optical properties and the coloration for CeSF in the ``{\it f-d} scenario'' by 
using the following procedure: starting from our LDA band structure described above, we apply
an artificial upward shift of 1.5 eV 
to the partially filled Ce 4$f$ band. This choice is
in contradiction to the results of our {\it GW} simulations, see Sec. of this supporting information, but
mimicks the situation of other ceria where a gap opens between $f$ and $d$ bands.
The resulting ``{\it f-d}'' electronic structure is depicted in Fig.~\ref{pshift1}.

In this case the occupied Ce $4f$ band is located well above the top of the S $3p$ band, hence the absorption edge is due to Ce $4f$-Ce 5$d$ ``intra-atomic'' transitions.
The calculated value of the band gap is 1.52 eV, which is smaller than the experimental one. Of course, the gap value depends on the choice of the shift.
However, the sharpness of absorption edge is not sensitive to this parameter.
One may notice, that, in agreement with previous electronic structure calculations \cite{Goubin2004}, the bottom of the conduction band is very dispersive. Therefore, one should expect a rather slow
growth of the joint density of states right above the gap.

\begin{figure*}
  \begin{center}
   \includegraphics[width=.9\textwidth]{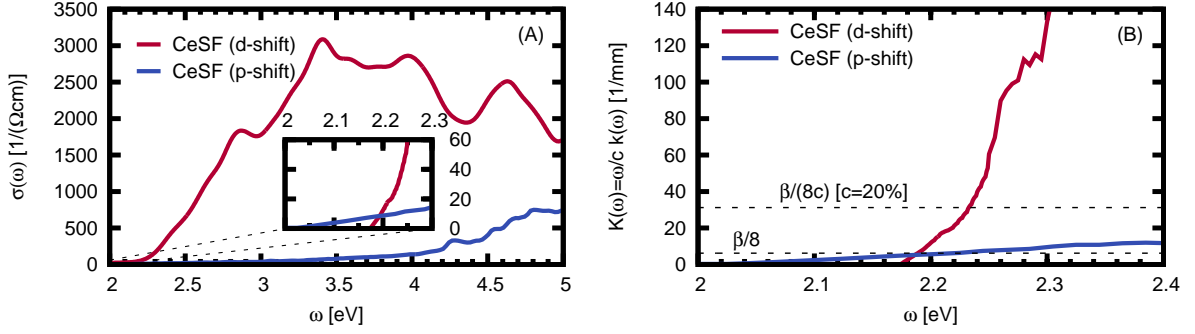}
      \caption{{\bf The two scenarios of CeSF~: impact on the optical conductivity and the absorption coefficient.} Comparison of the (true) ``{\it p-f}'' the (artificial) ``{\it f-d}'' scenario for (A) the optical conductivity and (B) the absorption coefficient. The quality measure $\beta/8c$ in the right panel uses again $\beta=50$/mm.}
      \label{pshift2}
      \end{center}
\end{figure*}

Here, we verify this qualitative analysis and  demonstrate that the ``{\it f-d}''
scenario fails to account for the established usability of CeSF as a high quality red pigment.
Shown in \fref{pshift2} (a) and (b) are the (polarization averaged) optical conductivities and absorption coefficients for both scenarios.
The gaps in both setups differ by several 100meV, explaining the different energies at which optical absorption
sets in. The other major difference in the optical spectra is that the intensity in the ``{\it f-d} scenario''
is significantly lower than for the ``{\it p-f}'' scenario suggested by our {\it GW} calculations (see above).
Indeed, the absorption coefficient $K(\omega)$ shown in \fref{pshift2} (b) fails to pass our quality measure for good pigments even in the bulk limit (concentration
$c=100\%$)~: $K(\omega)$ surpasses $\beta/8$ (with $\beta=50$/mm) at only about 0.2eV above the onset.

\begin{figure*}
  \begin{center}
   \includegraphics[angle=-90,width=.6\textwidth]{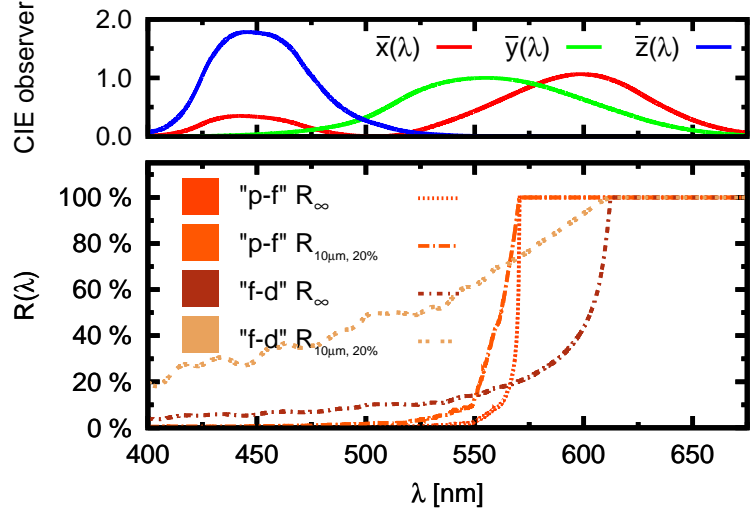}
	\caption{{\bf The theoretical diffuse reflectance.} Comparison of the diffuse reflectance $R(\lambda)$ and the corresponding colors in the ``{\it p-f}'' and the ``{\it f-d}'' scenario.}
      \label{pshift3}
      \end{center}
\end{figure*}

The consequences of this large variation in the sharpness of the absorption onset are displayed in \fref{pshift3}, which shows the diffuse reflectance and the resulting color
for both scenarios (the detailed color coordinates are in Table~1~: 
The bulk color (see $R_\infty$) is darker than the prominent orange/red of real CeSF that we reproduce in the ``{\it p-f}'' scenario. This mostly owes to the smaller value of the optical gap. As can be seen by looking at the red stimulus of the CIE observer, large parts of the red spectrum are cut off in the p-setup, decreasing the overall luminosity.  The shallowness of the onset of absorption, causes admixtures of 5-10\% of green and even blue, which slightly increases the brightness,
but in a rather uniform fashion, i.e.\ mostly adding components of gray.

\begin{table*}  
\begin{tabular}{|c|c|c|ccc|cc|}
 \hline
 & thickness & concentration& $\phantom{X}$X$\phantom{X}$ & $\phantom{X}$Y$\phantom{X}$ & $\phantom{X}$Z$\phantom{X}$ & $\phantom{X}$x$\phantom{X}$ & $\phantom{X}$y$\phantom{X}$\\
\hline
CeSF &$\infty$ & $100\%$             & 0.60  &  0.39  &  0.00 &0.60 &0.39 \\
``{\it p-f}'' scenario &$\infty$ & $\phantom{0}20\%$   & 0.63  &  0.44  &  0.01 &0.58 &0.41 \\
     &$10\mu$m & $\phantom{0}20\%$   & 0.65  &  0.47  &  0.01 &0.57 &0.42 \\
\hline
CeSF  &$\infty$ &$100\%$                    & 0.36  &  0.26  &  0.01 &0.50 &0.37 \\
``{\it f-d}'' scenario     &$\infty$ &$\phantom{0}20\%$    & 0.55  &  0.50  &  0.28 &0.42 &0.37 \\
     &$10\mu$m &$\phantom{0}20\%$           & 0.74  &  0.73  &  0.44 &0.39 &0.38 \\
\hline
HgS  &$\infty$ &$100\%$              & 0.53  &  0.33  &  0.01 &0.67 &0.33 \\
     &$\infty$ &$\phantom{0}20\%$    & 0.60  &  0.44  &  0.08 &0.54 &0.39 \\
     &$10\mu$m &$\phantom{0}20\%$    & 0.65  &  0.52  &  0.08 &0.52 &0.42 \\
     \hline
\end{tabular}
    \caption{{\bf The colors of CeSF within the (true) ``{\it p-f}'' and the (artificial) ``{\it f-d}'' scenario, and HgS.} Shown are the CIE 1931 tristimulus values XYZ and  the chromaticities xy (z=1-x-y)
    of both compounds, for different polarizations, pigment concentrations and sample thicknesses on a white substrate.}
\end{table*}

When using ``{\it f-d}'' CeSF as a realistic pigment, i.e.\ as a 20$\%$ admixture in a layer of 10$\mu$m on a white substrate, the color becomes a grayish ocher. This is due to the large increase in
contributions from all colors, as is evident from the diffuse reflectance in \fref{pshift3}.
As expected from our analysis (based on our quality measure for the absorption coefficient), this instability with regard to the deployment in a realistic setup, i.e.\
the dependence of the color on the concentration and the film thickness, make ``{\it f-d}'' CeSF unsuitable for practical applications.
This finding being at variance with the fact that CeSF is such a good pigments that it is indeed 
commercially used, thus strengthens our interpretation that in 
real CeSF the onset of absorption is due to $p$-$f$ transitions, as described by our
calculations.

\section{The color of matter}

The color of a pigment, or any object in general, owes to its diffuse, i.e. non-directional, reflectance.
This has to be contrasted to the specular reflectivity of geometric optics as given by Fresnel's formulae.
While the latter is entirely determined by the absorption properties of the system
(for an application to colors and correlated materials see Ref.~\cite{me_psik,optic_epl}), the 
diffuse reflectance requires the accounting for the scattering of light due to imperfections of the specimen,
such as impurities, grain boundaries or granularities.
These are, by definition, effects that are beyond a first principles description of extended, perfect solids.
We therefore resort to a modelisation for objects that both absorb and scatter light, where the latter
is accounted for by an effective scattering rate $\beta$.
While the latter in principle depends (like the absorption coefficient) on the frequency of the light, we shall demonstrate, for the case of CeSF, that this
dependence is weak and non-essential for the fidelity of the calculated color.

\begin{figure*}[!t]
        \centering
                \includegraphics[angle=-90,width=0.75\textwidth]{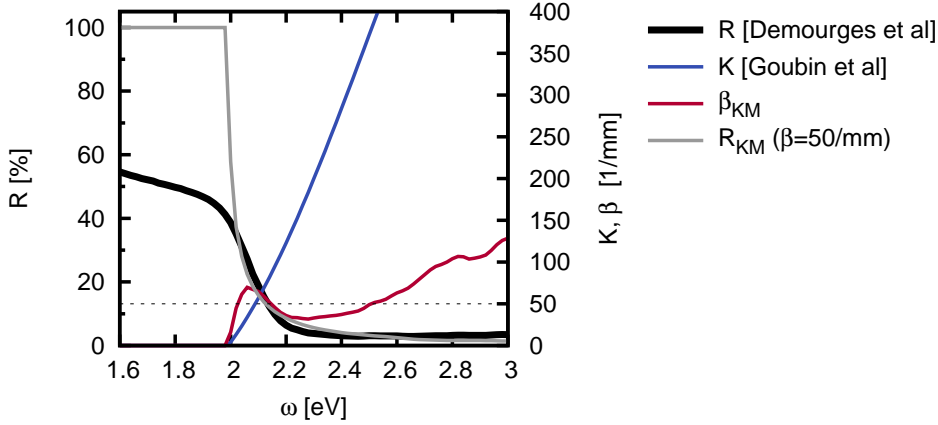}
        \caption{{\bf The diffuse reflectance.} Experimental diffuse reflectance $R$ of CeSF from Demourges et al.\cite{Demourgues2001},
          absorption coefficient $K=w/c\;\mathrm{Im}\sqrt{\hat{\epsilon}}$ from the dielectric function $\hat{\epsilon}$ of Goubin et al.\cite{Goubin2004}, and 
          scattering coefficient $\beta_{KM}(\omega)$, as extracted via the KM model. 
          Also shown is the KM diffuse reflectance $R_{KM}$ that results from the experimental absorption coefficient $K$ and an energy-independent
          scattering coefficient $\beta=50$/mm. The magnitude of the latter is indicated by the dashed line.}
        \label{figbeta}
\end{figure*}

\subsection{Effective medium theory for light scattering in inhomogeneous media -- the Kubelka-Munk model}

The model that we employ for our calculations is owing to Kubelka and Munk\cite{KM,Kubelka1948}, and is based on
considering two light fluxes, $J_+$ and $J_-$, that account for light propagating along a coordinate $x$ downwards or upwards in a film of thickness $L$ deposited on a substrate of reflectance $R_S$.

Using the absorption coefficient $K=\frac{\omega}{c}\mathrm{Im}\left(n+\im k\right)=\frac{\omega}{c}k$ and introducing the effective scattering parameter $\beta$, the
differential fluxes follow (for details see e.g.\ Refs.~\cite{TdC,Levinson2005})~:

\begin{eqnarray}
\frac{dJ_+(x)}{dx}&=&-2KJ_+(x)-\beta\left(J_+(x)+J_-(x) \right)\\
\frac{dJ_-(x)}{dx}&=&\phantom{-}2KJ_-(x)+\beta\left(J_+(x)+J_-(x) \right)
\end{eqnarray}

\noindent
From this, we obtain for the reflectance $R(x)=-J_+(x)/J_-(x)$ at depth x 

\begin{equation}
\frac{dR}{R}=\beta dx \left( R+\frac{1}{R}-2\left(1+\frac{2K}{\beta}\right) \right)
\end{equation}

\noindent
integrating $x$ from 0 to $L$, and thus $R$ from $R(0)=R_S$,  the substrate reflectance, to $R(L)$, and solving for $R(L)$ yields

\begin{eqnarray}
	R(L)=\frac{  R_\infty^{-1} (R_S-R_\infty)-R_\infty (R_S-R_\infty^{-1})e^{\beta L(R_\infty^{-1}-R_\infty)}    }
	{ (R_S-R_\infty)-(R_S-R_\infty^{-1})e^{\beta L(R_\infty^{-1}-R_\infty)}   } 
	\label{KML}
\end{eqnarray}

\noindent
where 

\begin{equation}
R_\infty=\alpha-\sqrt{\alpha^2-1}
\label{km1}
\end{equation}

\noindent
with $\alpha=1+2K/\beta$ corresponding to the reflectance of a semi-infinite sample (bulk limit).
Here, all quantities depend on the wavelength or frequency of the light.
Using a pigment as an admixture to a (mostly transparent) host material can be modeled, in the simplest possible ansatz,
by rescaling the absorption with the concentration $c$, i.e. using in the above equations $cK$ instead of $K$.

\subsection{The scattering parameter $\beta$}

The above description on the diffuse reflectance relies on the 
scattering parameter $\beta$ that is
not accessible from first principle calculations for the perfect solid.
In fact, $\beta$ is a property that depends on the microstructure of
the material.
It arises thus the question of how to determine typical values of
this parameter in practice, and, in particular,
whether its frequency dependence is crucial for the calculation of the actual color.

In the case of CeSF there exist experimental measurements of both, the
diffuse reflectance $R$\cite{Demourgues2001} and the complex refractive index $\hat{n}=n+\im k$\cite{Goubin2004}. 
Assuming the validity of the KM model, \eref{km1} can thus be inverted to extract the
scattering coefficient~:
$\beta_{KM}=\frac{2K}{\alpha-1}$, 
 with $\alpha$ expressed as $\alpha=\frac{1}{2}(R+\frac{1}{R})$.

Fig. \ref{figbeta} shows $R$, $K$, and the resulting $\beta_{KM}$ as a function of frequency.
As it turns out, the frequency dependence of the scattering coefficient $\beta$ 
is negligible. Indeed, approximating $\beta=50$/mm as indicated by the dashed line in \fref{figbeta},
the reflectance $R_{KM}$ within the KM model neatly traces the experimental data above the onset of absorption\footnote{Below the onset of absorption, the reflectance of the KM model is, by construction, always 100\%.
The difference to the experimental reflectance is owing to residual in-gap absorption, that, like the scattering coefficient,
is beyond an {\it ab inito} description of the perfect solid.
}.
This justifies our choice to use the approximate $\beta(\omega)=50\rm{mm}^{-1}$ in all of our
calculations.
This value is moreover consistent with the range of
values $\beta \approx 10-500 \rm{mm}^{-1}$ reported in the literature for a wide range of
industrial inorganic pigments~\cite{Levinson2005_2}.

\subsection{Opacity and covering power of thin films}

A brilliant color of the bulk, as resulting from $R_\infty$, does not guarantee that a compound can be used as a pigment.
While in some applications pigments are admixtures to the actual material the object is made of,
 pigments are mostly used in lacquer or paint. In other words it is not the bulk but the behavior
 in thin films, i.e. the opacity, that  defines a good coloring material.
Thus a good pigment must meet two requirements~:

\noindent
\paragraph{Imposing the pigment's color.}
 The worst case is a ``black'' substrate : 
 $R_S=0$, while $R_\infty\ne 0$.
Then it follows from \eref{KML}
\begin{equation}
	R(L)=R_\infty\frac{e^{\beta L(R_\infty^{-1}-R_\infty)}-1}{e^{\beta L(R_\infty^{-1}-R_\infty)}-R_\infty^2}\leq R_\infty
	\label{eq:}
\end{equation}
As a matter of fact the finite thickness of the film results, for the current case, in a more sharply defined absorption~: While
 $R(L)=1$ for $R_\infty=1$, $R(L)<R_\infty$ for $R_\infty<1$. 

\medskip

\noindent
\paragraph{Hiding the color of the substrate.}
Now the worst case is a ``white'' substrate\ : 
$R_S=1$, while $R_\infty\ll 1$.
Then
\begin{equation}
	R(L)=\frac{1+R_\infty e^{\beta L(R_\infty^{-1}-R_\infty)}}
	{R_\infty+e^{\beta L(R_\infty^{-1}-R_\infty)}}
	\label{Rlrs1} 
\end{equation}
In that case $R(L)\ge R_\infty$, which means that a finite film will have a broader tail in the reflectance and admixtures of wavelengths away from the absorption
edge are larger than for the bulk material. Hence the quality measure of a pigment that we introduce in the main manuscript considers a white substrate.

\subsection{From the diffuse reflectance to the perceived color\label{color}}

While many spectroscopic devices have been devised to analyse matter,
the spectrometer most commonly used is the human eye~: it sensors the response
to incident light in a range from roughly $\lambda=400$ to 700~nm, 
and the information gathered by the rods and cones of the retina are post-processed by our brain into a brightness and color sensation.

The calibration, or {\it  color matching functions},  of the three types of cones, responsible for the color perception, can be empirically measured~\cite{rgb1,rgb2}
 and used to setup a three-dimensional color space. Many variants of color schemes exist, and they are basically related by a unitary transformations of their coordinates.
In the present work, we employ the CIE 1931 XYZ color space~\cite{cie,1475-4878-33-3-301}. The corresponding matching functions, probing the response to red, green, and blue light,
are denoted $\bar{x} (\lambda )$, $\bar{y} (\lambda )$
and  $\bar{z} (\lambda )$, and are displayed in the top panel of Fig.~4 of the manuscript or, here, \fref{pshift3}. 
 
Knowing the reflectivity $R(\lambda)$ of a material or object (as e.g. in the bottom panel of Fig.~4 
of the main text or, here, \fref{pshift3}), and the spectral distribution $S(\lambda)$ of the light source, i.e.\ accounting for the type of light that the considered object is illuminated with, the XYZ color coordinates, or {\it tristimulus values}, are calculated as (see e.g. Ref.~\cite{color})
\begin{eqnarray}
\biggl( 
X, Y, Z
\biggr)&=&k\int \dif\lambda\, R(\lambda) S(\lambda) \biggl( 
\bar{x} (\lambda ), \bar{y} (\lambda ), \bar{z} (\lambda )
\biggr)
\end{eqnarray}
where $k$ is a normalization constant chosen such that $k\int\dif\lambda S(\lambda)\bar{y}(\lambda)=1$.
As light source we use in our calculations the CIE standard illuminant D65~\cite{D65}, which is appropriate for usual day light.
While the tristimulus coordinates include information on both color and brightness, in color pictograms
often only the color information given by the normalized {\it chromaticity values}
\begin{eqnarray}
x&=&\frac{X}{X+Y+Z}\nonumber\\
y&=&\frac{Y}{X+Y+Z}\\
z&=&\frac{Z}{X+Y+Z}\nonumber
\end{eqnarray}
which are shown in Table~1 for the two scenarios of CeSF, as well as for HgS.

\section{Interband Absorption in Two-band Model}\label{2b_model_suppl}

\begin{figure*}
\begin{center}
\includegraphics[width=0.65\textwidth]{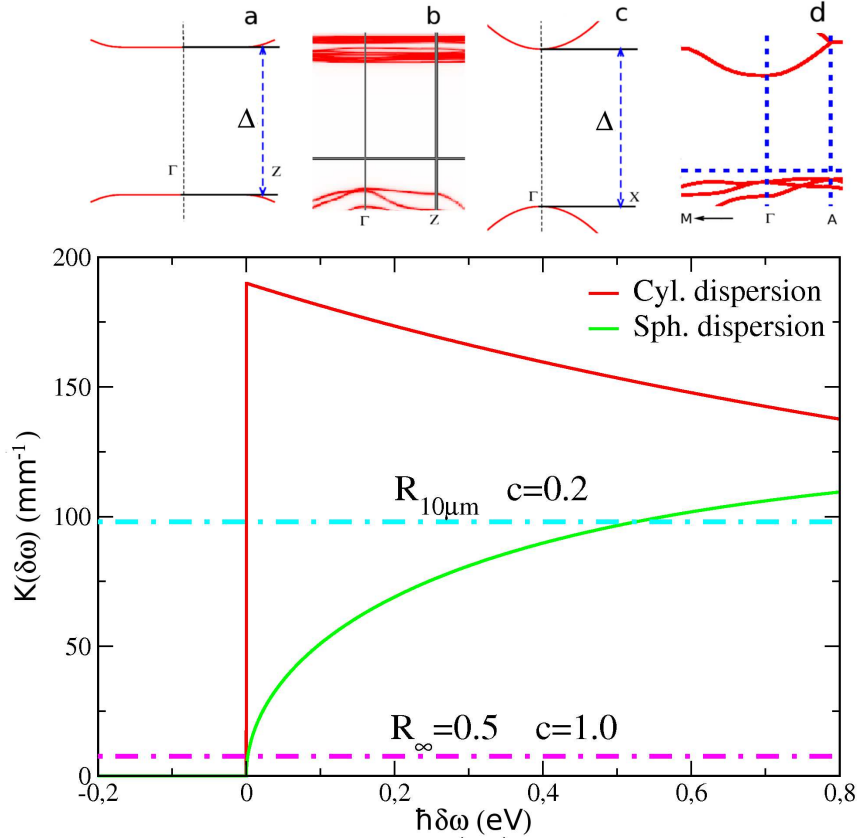}
\caption{Upper panel: a. Two model bands ``cylindrical'' along the $\Gamma-Z$ path (along other two paths $\Gamma-X$ and $\Gamma-Y$ they have the same shape as shown in c).; b. Zoom at the CeSF bands along the $\Gamma-Z$ path.;  c. Two model bands in the spherical case; d. Zoom at the HgS bands along the $M-\Gamma-A$ path.
Lower panel: Absorption $K(\delta\omega)$ for the two-band models with spherical and cylindrical band dispersions. The cyan and pink horizontal lines
indicate the threshold value of $K$ for the semi-infinite bulk and thin film cases, see the text}\label{K_model}
\end{center}
\end{figure*}

In order to understand how the topology of the bottom of conduction/top of valence bands in CeSF influences its optical properties
we consider a simplified model, in which we include a single valence and a single conduction band with analytical dispersion laws. This model is adequate to describe
the optical conductivity in the vicinity of the absorption edge, where the relevant interband transitions are between single valence and conduction
bands, which have a quasi-two-dimensional shape in the case of CeSF (see Fig.~\ref{K_model}b). Thus, to model the absorption edge in CeSF we consider the case of ``cylindrical'' valence and conduction
bands (Fig.~\ref{K_model}a). Three-dimensional (``spherical'') bands (Fig.~\ref{K_model}c) provide a more adequate model of the actual states involved in optical transitions at the absorption edge of HgS
(especially for the conduction band, see Fig.~\ref{K_model}d).

In calculations of the interband absorption one starts from the real part of the optical conductivity (Eq.~1 in the main text), which in the case of zero temperature can be written as:
\beq
\mathrm{Re}\sigma(\nu)=\frac{2\pi e^2 \hbar}{V}\sum_{\vk}\int\frac{d\omega}{\nu}tr[v(\vk)A(\vk,\omega)v(\vk)A(\vk,\omega+\nu)].
\label{opt_cond}
\eeq
The imaginary part of the dielectric function (in CGS units) and absorption $K$ are then evaluated from $\mathrm{Re}\sigma(\omega)$ as 
\beq
\mathrm{Im}\epsilon^{cgs}(\omega)=\frac{4\pi}{\omega}\mathrm{Re}\sigma^{cgs}(\omega)=\frac{\mathrm{Re}\sigma(\omega)}{\omega \epsilon_0}
\eeq
\beq
K(\omega)=\frac{\omega k(\omega)}{c}=\frac{\omega \mathrm{Im}\epsilon^{cgs}(\omega)}{2\hat{n}c}=\frac{\mathrm{Re}\sigma(\omega)}{2\hat{n}c\epsilon_0},
\label{abs2}
\eeq
where $\hat{n}=n+\im k$ is the refractive index, $c$ is the speed of light, $\epsilon_0$ is the vacuum permittivity. The absorption $K(\omega)$ has the dimension of inverse length.

Neglecting the $\vk$-dependence for the velocities, $v_{12}(\vk)v_{21}(\vk)=-v^2$ one may carry out the energy integration in (\ref{opt_cond}) and obtain the optical conductivity expressed through the ``joint density of states'':
\begin{eqnarray}
\int d\omega \delta\left[\hbar\omega-E_v(\vk)\right] \delta\left[\hbar\omega-E_c(\vk)+\hbar\nu\right]\qquad\nonumber\\
=
\frac{1}{\hbar}\delta\left[E_v(\vk)-E_c(\vk)+\hbar\nu\right],
\end{eqnarray}
where $E_v(\vk)$ and $E_v(\vk)$ are the valence and conduction bands dispersions, respectively. Then (\ref{opt_cond}) reads as follows: 
\beq
\mathrm{Re}\sigma(\nu)=-\frac{2\pi e^2 \hbar }{\nu}v^2 \int \frac{2d^3k}{8\pi^3}\delta\left[E_v(\vk)-E_c(\vk)+\hbar\nu \right].
\label{opt_cond_2}
\eeq
Let us now introduce the band dispersions corresponding to the ``spherical'' (Fig.~\ref{K_model}c)
\beq
E_v(\vk)=-\frac{\hbar^2 k^2}{2 m_v},
\eeq
\beq
E_c(\vk)=\frac{\hbar^2 k^2}{2 m_c}+\Delta,
\eeq
and ``cylindrical'' (Fig.~\ref{K_model}a)
\beq
E_v(\vk)=-\frac{\hbar^2 (k_x^2+k_y^2)}{2 m_v},
\eeq
\beq
E_c(\vk)=\frac{\hbar^2 (k_x^2+k_y^2)}{2 m_c}+\Delta,
\eeq 
cases. Here, $m_{v(c)}$ are the valence (conduction) band's effective masses and $\Delta$ is the band gap value.

By substituting the $d^3k$ integration in (\ref{opt_cond_2}) with the integration over the constant energy surface $\int d^3k \delta[E(\vk)-E_0]... \rightarrow \int \frac{dS}{\nabla_k[E(\vk)-E_0]}...$ one obtains the real part of the optical conductivity:
\beq
\mathrm{Re}\sigma^{sph}(\delta\omega)=\frac{2 \sqrt{2} e^2 v^2}{(\Delta+\hbar\delta\omega)\pi \hbar^2} m_r^{3/2} \left(\hbar\delta\omega\right)^{1/2},
\eeq
\beq
\mathrm{Re}\sigma^{cyl}(\delta\omega)=\frac{2 e^2v^2}{(\Delta+\hbar\delta\omega) a \hbar} m_r,
\eeq
for the spherical and cylindrical dispersions, respectively. Here, $m_r$ is the reduced effective mass $m_r=m_c m_v/(m_c+m_v)$. 
$\hbar\delta\omega$ is the photon energy relative to the absorption edge, $\hbar\nu=\Delta+\hbar\delta\omega$. Of course, $\mathrm{Re}\sigma(\delta\omega)=0$ within the gap ($\delta\omega <0$), hence one may immediately see that $\mathrm{Re}\sigma^{cyl}(\delta\omega)$ exhibit a jump at the absorption edge, while $\mathrm{Re}\sigma^{sph}(\delta\omega)$ is continuous everywhere.

The corresponding absorptions (\ref{abs2}) for $\delta\omega \geq 0$ (for $\delta\omega < 0$ the absorption is again zero)  read 
\beq
K_{sph}(\delta\omega)=\frac{\sqrt{2} e^2 v^2}{\pi(\Delta+\hbar\delta\omega) \hat{n}\hbar^2 c \epsilon_0}m_r^{3/2} \left(\hbar\delta\omega\right)^{1/2}
\label{abs_sphere}
\eeq
\beq
K_{cyl}(\delta\omega)=\frac{ e^2v^2}{(\Delta+\hbar\delta\omega) a \hbar \hat{n} c \epsilon_0} m_r .
\label{K_cyl}
\eeq

To model the CeSF case one may assume a flat conduction band (the dispersion of the Ce 4$f$ states at the bottom of conduction band is negligible, see Fig.~\ref{K_model}b), 
hence $m_r=m_v$. For the effective mass we used the value $m_r=m_v$ equal to $0.6 m_0$, which was extracted 
from the curvature of the topmost S 3$p$ band at $\Gamma$ in the CeSF LDA+DMFT momentum-resolved spectral function, Fig.~\ref{CeSF_bands_full}. Other parameters entering in 
Eqs.~(\ref{abs_sphere},\ref{K_cyl}) were chosen as follows: $\Delta$=2.1 eV, $\hat{n} \approx 2$ was obtained from the full complex 
optical conductivity of CeSF, the average Fermi velocity can be estimated to be $v=V_{pf}\cdot\delta_{Ce-S} \sim 0.3$~eV\AA, where $\delta_{Ce-S} \approx$ 3\AA\ 
is the distance between next-nearest Ce and S sites in CeSF, and $V_{pf}$ is a hybridization matrix element between the involved orbitals.

The resulting absorptions for both the spherical and cylindrical cases are displayed in Fig.~\ref{K_model} with the same set of model parameters 
 employed for both cases. One may see that the stepwise behavior of the curve in the ``cylindrical'' case results in a much higher absorption just 
above the edge  as compared to the spherical case. This leads to an important difference in the color properties in the two cases. To quantify those properties we again 
employ the KM model with the scattering parameter $\beta=50$/mm and compute the characteristic absorption thresholds for the two cases considered in the article:
the semi-infinite bulk of pure pigment and a thin layer of 10 $\mu$m
with the pigment concentration $c=$0.20.
In the ``spherical'' case the characteristic absorption threshold $\delta\omega_{1/2}$, for which the diffuse reflectance 
$R=1/2$, is equal to $\sim$7 meV for the semi-infinite bulk. Hence, the absorption edge is sufficiently sharp in this case. However, in the case of a thin film we obtain $\delta\omega_{1/2}$=0.52~eV,
which is substantially larger than the characteristic frequency interval
$\delta\omega_c \approx$0.2eV (see the main text).
 Therefore, 
the ``model material'' with the spherical bands (Fig.~\ref{K_model}c) would hardly make a very good pigment. In contrast, in the model system 
with the two-dimensional band topology  the diffuse reflectance will undergo a discontinuous transition from $R=1$ to to $R\approx 0$ exactly at the edge. Hence,
 a quasi two-dimensional topology of the conduction/valence bands naturally leads to a sharp onset of absorption and drop 
in $R$ as predicted by the LDA+DMFT simulations for CeSF (see Fig.~4 in the main manuscript).


\end{document}